# The VLT-FLAMES Survey of Massive Stars

Chris Evans[1], Ian Hunter[2], Stephen Smartt[2], Danny Lennon[3,8], Alex de Koter[4], Rohied Mokiem[5], Carrie Trundle[2], Philip Dufton[2], Robert Ryans[2], Joachim Puls[6], Jorick Vink[7], Artemio Herrero[8], Sergio Simón-Díaz[9], Norbert Langer[10], Ines Brott[10]

[1] UK ATC, Edinburgh, UK
[2] Queen's University Belfast, Northern Ireland, UK
[3] Space Telescope Science Institute, Baltimore, USA
[4] University of Amsterdam, The Netherlands
[5] OC&C Strategy Consultants, Rotterdam, The Netherlands
[6] Universitäts-Sternwarte, Munich, Germany
[7] Armagh Observatory, Northern Ireland, UK
[8] Institutio d'Astrofísico Canarias, Tenerife, Spain
[9] Geneva Observatory, Switzerland
[10] University of Utrecht, The Netherlands

**The VLT-FLAMES Survey of Massive Stars was an ESO Large Programme to understand rotational mixing and stellar mass-loss in different metallicity environments, in order to better constrain massive star evolution. We gathered high-quality spectra of over 800 stars in the Galaxy and Magellanic Clouds. A sample of this size is unprecedented, enabled by the first high-resolution, wide-field, multi-object spectrograph on an 8-m telescope. We developed spectral analysis techniques that, in combination with non-LTE, line-blanketed model atmospheres, were used to quantitatively characterize every star. The large sample, combined with the theoretical developments, has produced exciting new insights into the evolution of the most massive stars.**

Massive stars dominate their local environment via their intense radiation fields, their strong winds and, ultimately, in their death as core-collapse supernovae. Larger telescopes and new instrumentation have provided the means to observe individual massive stars beyond the Milky Way – in the Large and Small Magellanic Clouds (LMC and SMC), in M31, and beyond. In parallel to this, the theoretical models needed to interpret the observations have become increasingly sophisticated, incorporating the effects of stellar winds (a far from trivial problem!) and opacities for the millions of metallic transitions occurring in their atmospheres. While our understanding of massive stars has improved significantly over the past 30 years, key questions remain concerning the role of metallicity (i.e. environment) on their stellar winds and rotational velocities, and the efficiency of rotational mixing in their interiors and atmospheres.

The delivery of FLAMES to the VLT was the catalyst for our Large Programme, targeting O- and early B-type stars in fields centred on stellar clusters in the Galaxy and in the Magellanic Clouds (e.g. NGC346 in the SMC, Figure 1). The LMC and SMC are metal poor when compared to the solar neighbourhood, with metallicities of ~50% and 25% solar, respectively. With the multiplex advantage of FLAMES, we were able to obtain a large observational sample of massive stars, in three distinctly different environments. Six of the standard, high-resolution (R ~ 20,000) settings of the Giraffe spectrograph were used, giving continuous coverage from 385-475 nm in the blue, combined with red spectra covering 638-662 nm (which includes the Hα Balmer line). An overview of the observations was reported in this publication by Evans et al. (2005a), with more detailed descriptions given by Evans et al. (2005b; 2006). This unique dataset has enabled us to test theoretical predictions of the physical properties of massive stars, and to provide valuable empirical information to groups working on evolutionary models. Here we summarise the key results from the ten refereed papers now published from the survey.

### Metallicity-dependent stellar winds

The out-flowing winds observed in massive stars are thought to be driven by momentum transferred from the radiation field to metallic ions in their extended atmospheres. A logical consequence of this mechanism is that the intensity of the outflows should vary with metallicity ($Z$), with the prediction from Monte-Carlo models that the mass-loss rates should scale as $Z^{0.69}$ (Vink et al., 2001). Such predictions are far from just an interesting quirk of stellar astrophysics; reduced mass-loss rates at low metallicity mean that an O-type star will lose less of its initial mass and angular momentum over its lifetime – this not only has a direct effect on the late stages of stellar evolution, but also on the nature of the final explosion as a supernova or a gamma-ray burst (GRB).

Analysis of O-type spectra with model atmospheres can be a complex, time-consuming process. In addition to the usual parameters used to characterise a star (temperature, luminosity, gravity, chemical abundances), we also need to describe the velocity structure and mass-loss rate of the wind. For the FLAMES project we adopted an innovative, semi-automated approach to the analysis, employing genetic algorithms to fit the observations with synthetic spectra from FASTWIND model atmospheres (Mokiem et al. 2005). Comparisons with published results for Galactic stars demonstrated the validity of the method, which was then used to analyse the O-type spectra from the FLAMES observations in the SMC and LMC (Mokiem et al., 2006; 2007a).

To investigate the effects of metallicity we have considered the modified wind-momentum–luminosity relation (WLR). This is a function of the mass-loss rate, terminal velocity and stellar radius, which is well correlated with stellar luminosity. In Figure 2 we show the observed WLRs for the LMC and SMC samples, compared to Galactic results obtained using the same models; the 1-sigma confidence intervals are shown as grey areas. The three empirical fits are clearly separated, providing quantitative evidence for reduced wind-intensities at decreased metallicities, and showing for the first time that the wind-intensities of stars in the LMC are intermediate to those in the Galaxy and SMC. Figure 2 also shows the theoretical predictions using the prescription from Vink et al. (2001). There is a systematic offset between the observed and predicted relations (perhaps arising from clumping of material in the winds), but the relative separations are in good agreement. From the FLAMES results we find a $Z$-dependence with exponents in the range 0.72 to 0.83 (depending on assumptions regarding the clumping), as compared to $Z^{0.69\pm0.10}$ from theory (Vink et al., 2001).

This observational test is important for a number of areas in contemporary astrophysics. The reduced mass-loss rates at lower metallicity mean that less angular momentum will be lost over the star's lifetime, i.e. an evolved star in a low metallicity environment would be expected to retain a larger fraction of its initial rotational velocity compared to a similar star in the Milky Way. Indeed, the rotational velocity distribution for our unevolved (i.e. luminosity class IV or V) SMC stars, appears to have preferentially faster velocities when compared to Galactic results – unfortunately the statistical significance of this result is limited by the relatively small number of unevolved O-type stars in our sample, but we will return to this later using the much larger sample of B-type stars. These effects mean that at low metallicity a larger fraction of stars would be expected to undergo chemically-homogeneous evolution, suggested as a channel for the progenitors of long-duration GRBs (e.g. Yoon et al., 2006).

This empirical test of the mass-loss scaling also reinforces the need to consider metallicity when interpreting observations of distant, unresolved star-forming galaxies, e.g. via the inclusion of low-metallicity spectral libraries in population synthesis codes to interpret the rest-frame ultraviolet observations of Lyman-break galaxies.

**Chemical composition of the Magellanic Clouds**

Studies of stellar abundances in rapidly-rotating stars are complicated by their broadened lines, which is why most observational effort in the past has been directed at narrow-lined (i.e. slowly-rotating) stars. Thus, before investigating the global trends in the whole sample, we first used the narrow-lined B-type stars (vsin$i$ < 100 km/s) to determine precise, present-day abundances for the LMC and SMC.

The TLUSTY model atmosphere code was used to analyse this sample of over 100 B-type stars (Hunter et al., 2007; Trundle et al., 2007). The present day composition of the LMC and SMC, as traced by these slowly-rotating B-type stars, is listed in Table 1. Note that the relative fraction of our SMC and LMC abundances, as compared to solar values, changes from element to element. Specifically, it has been known for some time that the initial abundances of carbon and nitrogen are significantly underabundant when compared to the heavier elements in the Clouds, i.e. simply scaling solar abundances does not best reproduce the observed patterns.

**Stellar temperatures as a function of metallicity**

The narrow-lined B-type stars were also used to investigate effective temperatures as a function of spectral type (Trundle et al. 2007); the resulting temperature calibrations are presented in Table 2. The well-known dependence of temperatures on luminosity class is evident, i.e. supergiant stars with their lower gravities, and more extended atmospheres, are found to be cooler than dwarfs of the same spectral type. We also find evidence of a metallicity dependence of the temperatures at a given spectral type. This is thought to arise from the effects of line blanketing, whereby the cumulative opacity of the huge number of spectral lines introduces additional back scattering, leading to changes in the ionization balance and effective temperature in the atmosphere (see Mokiem et al. 2006, and references therein). This effect is well documented in O-type stars (e.g. Mokiem et al., 2007a), but the FLAMES survey has provided the first evidence for it in B-type stars – there is a relatively small difference between the results for the LMC and SMC, but there is a clear offset seen for the Galactic stars.

Calibrations such as these are widely used to provide temperature estimates in instances where high-quality spectroscopy of a star is not available, but its spectral type is known; the FLAMES results highlight the need, and provide the necessary information, to consider metallicity effects when adopting such temperature estimates.

**Low-metallicity stars spin faster**

The prevailing viewpoint for the past decade has been that rotation strongly influences the evolutionary path of O- and B-type stars. Furthermore, it has long been assumed that stars should rotate more quickly at low metallicities. While there has been some reasons to believe this (e.g. higher fractions of Be-type stars in the Clouds) it has never been verified quantitatively. As rotating stellar models predict that excess nitrogen and helium, produced during core hydrogen burning, can be mixed to the surface, abundances of these elements from the FLAMES survey can be used to test the theories.

We developed new spectral-analysis tools based on TLUSTY model atmospheres to rapidly analyse large numbers of quickly-rotating stars. We were able to determine physical parameters, rotational velocities and nitrogen abundances for all of the B-type stars observed in the LMC and SMC with velocities up to 300 km/s (~400 stars, Hunter et al., 2008a). The size of the sample is the most extensive to date, and the first in the Clouds that is large enough to model the underlying distribution of rotational velocities by assuming random angles of inclination of the rotation axes. As mentioned earlier, the O-type stars will be expected to slow down over their main-sequence lifetimes as they will lose angular momentum as a consequence of mass-loss by their winds; we therefore only considered stars with masses less than 25 $M_{sun}$. In Figure 3 we show the cumulative probability functions for vsin$i$ of the core-hydrogen-burning (i.e. giant and dwarf) B-type stars in the SMC and LMC – there is a clear difference between the two curves, with the SMC stars characterised by faster rotational velocities.

To extend this comparison to higher metallicity, we first needed to define an appropriate Galactic sample. Most of the Galactic stars observed in the survey were members of the central clusters, whereas our LMC and SMC stars are predominantly field stars. This distinction is important given that rotational velocities for our Galactic stars were found to be larger than for the field star population (Dufton et al., 2006). The Galactic curve shown in Figure 3 was therefore constructed using vsin$i$ results from published surveys of field stars (see further discussion by Hunter et al., 2008a). Assuming random angles of inclination, the median intrinsic rotational velocities for the Galactic, LMC and SMC stars (M < 25 $M_{sun}$) are 125, 135, and 183 km/s respectively. We have clear evidence (which is significant at the 3-sigma level) that the massive stars in the SMC rotate more quickly than in the Milky Way, and for the first time have a reliable intrinsic rotational velocity distribution in the SMC and LMC. The results for the O-type SMC stars have already been used by Yoon et al. (2006) to predict the rate of GRBs in the Universe from homogeneously-mixed massive stars. However, it is clear from recent work that stars in bound clusters appear to rotate significantly more quickly than stars in the field. It remains to be seen if the place of birth is as important as initial metallicity in determining the intrinsic rotation rate of a star. This is an important open question for future surveys of massive stars.

**Rotational mixing is not as dominant as we thought**

To investigate the impact of rotation on surface nitrogen abundances, new evolutionary models were calculated at LMC metallicity (Brott et al., in prep.). Aside from the effects of mass-loss, other factors lead to changes in vsin$i$ with time, primarily the contraction/expansion of the star, meridional circulation, internal magnetic fields and diffusion effects. Rather than simply scaling solar abundances, the new models adopt the chemical composition from Table 1. The mixing efficiency in the models was then calibrated to reproduce the observed surface nitrogen abundance at the end of core hydrogen burning for a 13 $M_{sun}$ model (the mean mass of the LMC stars in our sample).

Figure 4 shows the nitrogen abundances, as a function of vsin$i$, for the LMC B-type stars (Hunter et al., 2008b). Typical uncertainties are of order 0.2-0.3 dex, so the scatter in the results indicates genuine differences in the surface nitrogen enrichment in both the core-hydrogen-burning stars (dwarfs and giants, left panel) and the supergiants (right panel).

There are two groups (labelled as Groups 1 and 2 in the shaded regions of the left-hand panel) that appear inconsistent with rotating models. The blue points in Group 1 comprise rapidly-rotating stars that appear to have undergone little chemical mixing, and yet they have surface gravities that indicate they are near the end of core hydrogen burning. According to the single-star models that include the effects of rotational mixing, the observed nitrogen over-abundances in these stars are expected to be larger by ~0.5 dex (at least for the more massive objects, in which the mixing is expected to be most efficient).

We see no evidence of binarity in the spectra of many of these stars (although we note that the observations were not optimized for binary detection), presenting a conflicting picture when compared with the single-star predictions of rotationally-induced mixing.

The 14 (apparently single) core-hydrogen-burning stars in Group 2 are equally puzzling in that they are rotating very slowly (less than 50 km/s) and yet show significant nitrogen enrichment.  For a random orientation, we would expect about two of these to be rapidly-rotating stars viewed pole-on, but this is highly unlikely for all 14, and we conclude that the majority are intrinsically slow rotators.  Recent studies of Galactic ß-Cepheid stars have found a correlation between nitrogen enrichment and magnetic fields (Morel et al., 2006); perhaps the enrichments found in the slowly-rotating B-type stars in Figure 4 are somehow linked to magnetic fields.

The results for the supergiants can be considered as two groups: Group 3, with relatively normal levels of enrichment, and Group 4, with much larger abundances (12+log[N/H] > 7.6).  Simplistically one might think of these as pre-red-supergiant stars (Group 3) and post-red-supergiant stars (Group 4).  However, while the abundances in Group 4 are consistent with predictions, the models cannot reproduce their effective temperature on the Hertzsprung-Russell diagram; some of the enriched objects show evidence of binarity, so mass-transfer may also be important.  These results are also supported by analysis of the SMC and Galactic stars (Hunter et al., in prep.); reconciling these observations with the evolutionary models demands further study.

**A serendipitous benefit of multi-epoch service observations**

Owing to the time-sampling of the service observations, the survey has discovered a wealth of new binary systems, some of which will be the subject of forthcoming papers.  Moreover, cross-correlation of the radial velocities of the Calcium *K* line in repeat exposures of the same stars show a typical scatter of ±2 km/s, demonstrating the excellent stability of the Giraffe spectrograph.  The global picture in terms of binarity is also of interest – the lower limit to the binary fractions in our three fields with the best time coverage (N11, NGC346, and NGC2004) are in the range 25-35%.  Curiously, we find a much lower binary fraction for stars in the NGC330 FLAMES field (4%) – whether this is simply a consequence of less thorough time-sampling, or if the binary fraction is genuinely different to that found in the other fields remains unclear.  Optimised follow-up of each of the fields will provide more rigorous binary fractions, a vital constraint to models of star formation that is lacking in the current literature.

**Unanswered questions and problems**

The FLAMES survey has provided a valuable and unique source of empirical information, enabling a huge step-forward in our understanding of massive star evolution.  However, it has unexpectedly raised new and critical problems that still challenge our understanding of these enigmatic stars:

- While rotational mixing appears to play a role in the enrichment of surface nitrogen in massive stars, our results from B-type stars demonstrate that it is not the only process, particularly at low rotational velocities.  This result presents a significant new challenge to theorists working on evolutionary models.

- We have found tentative evidence that O-type stars in the SMC lose less angular momentum via their stellar winds than Galactic stars, i.e. the unevolved SMC stars are rotating more quickly.  This principle underpins one of the potential channels for long-duration GRBs, but the significance of our result is limited by the number of O-type stars in the FLAMES sample.  A more expansive programme examining the rotational velocities of O-type stars in the SMC is required to confirm this result.

- The intrinsic rotational velocity distribution of O- and B-type stars appear indistinguishable, but the B-type stars tend to rotate at a greater fraction of critical (Keplerian) velocity, potentially leading to a greater number of GRBs at low metallicity than predicted by current models – new theoretical calculations of GRBs from stars with initial masses ranging from 10 to 25 $M_{sun}$ are required to investigate this.

- Evolutionary models of single stars do not reproduce the observed temperatures of the nitrogen-rich B-type supergiants (Group 4 in Figure 4), i.e. they do not predict 'blue loops' at sufficiently high temperatures, nor high enough masses. Continued spectroscopic monitoring of the nitrogen-enhanced supergiants for long-period binaries would provide an essential constraint on further work in this area.

- There is compelling evidence that stellar winds in O-type stars are clumped. If the clump properties do not depend on metal content, nor the rate of mass-loss, the wind scalings presented here will not be affected. However, to quantify the effects of wind clumping properly requires further observational and theoretical investigation.

Finally, we note that all of the spectra from the survey are now publicly available at: "http://star.pst.qub.ac.uk/~sjs/flames/"

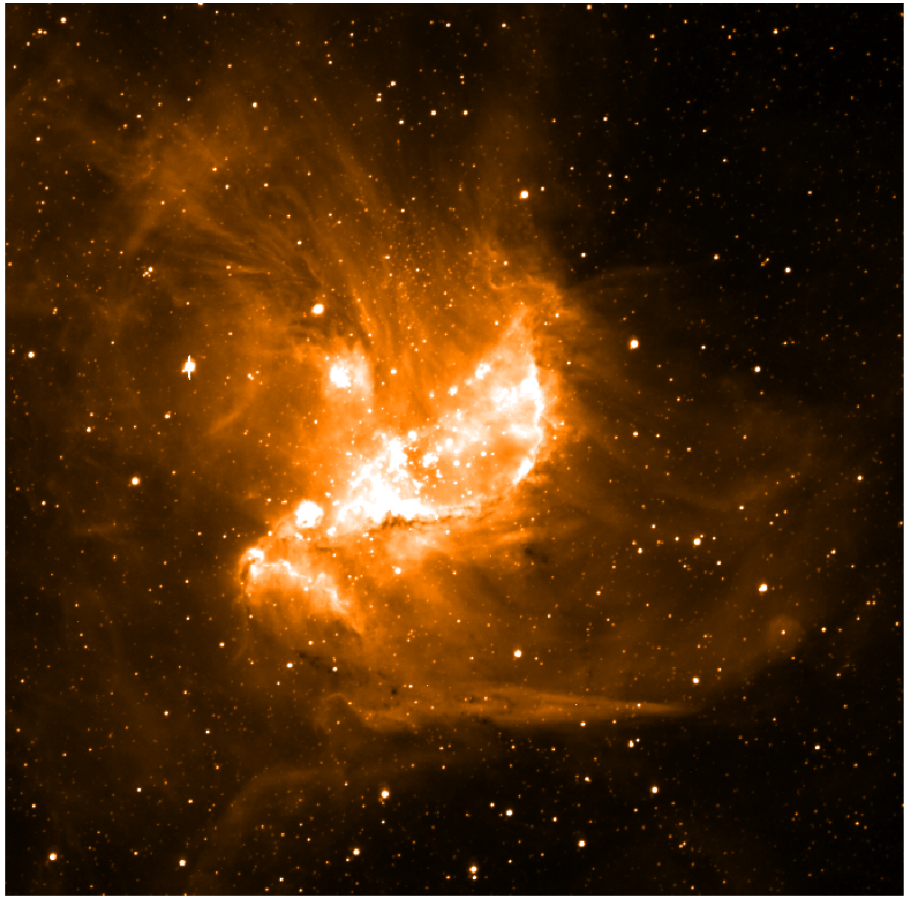

Figure 1: VLT-FORS Hα-image of NGC 346, the largest HII region in the Small Magellanic Cloud and one our target fields observed with FLAMES (E. Tolstoy/ESO Archive).

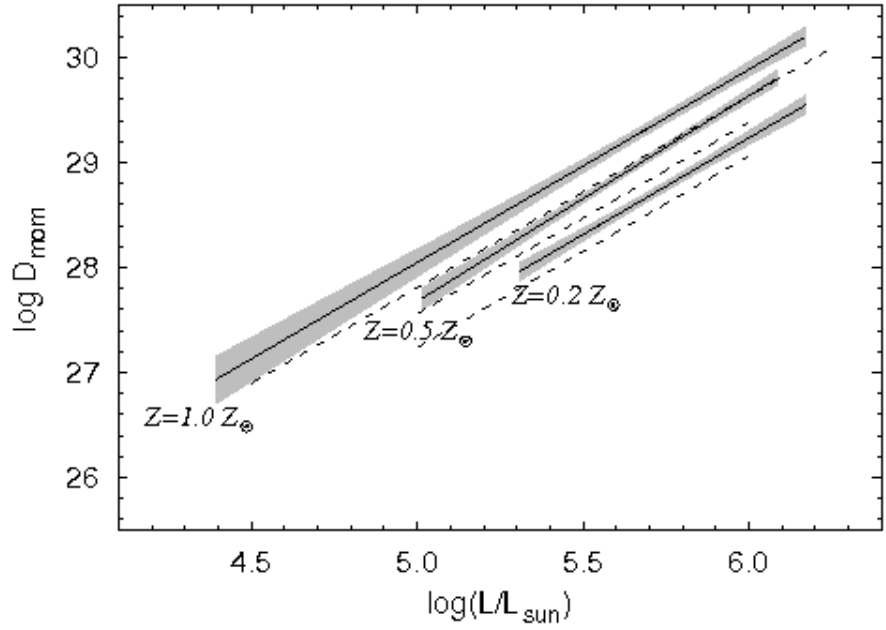

Figure 2: Comparison of the observed wind-momentum–luminosity relations (solid lines) with theoretical predictions (dashed lines). For each set, the upper, middle and lower relations correspond to Galactic, LMC and SMC results respectively.

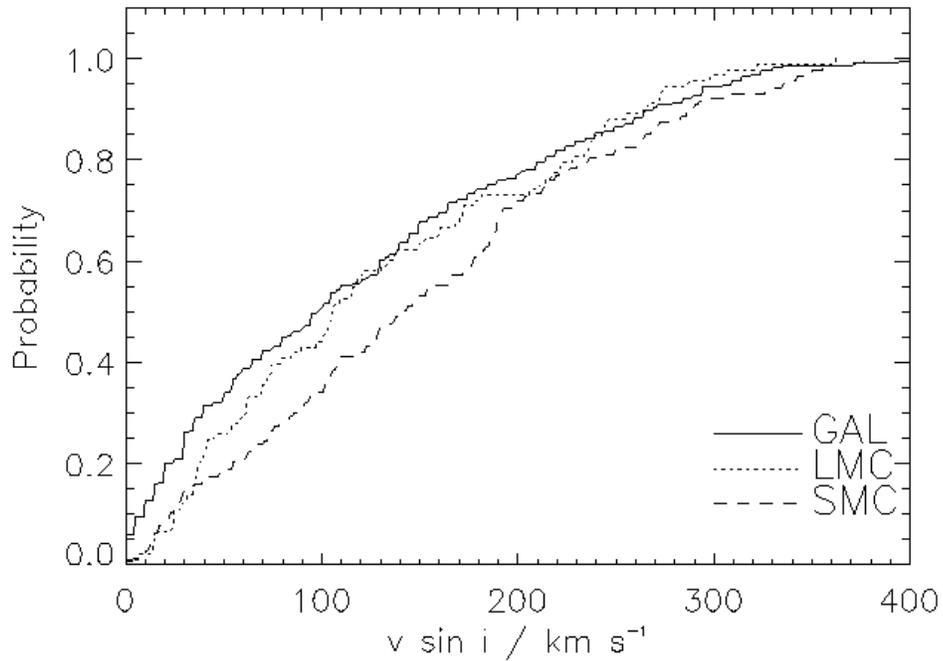

Figure 3: Cumulative distribution functions for the rotational velocities of Galactic field stars, compared with the LMC and SMC results from FLAMES – faster velocities are seen at lower metallicity.

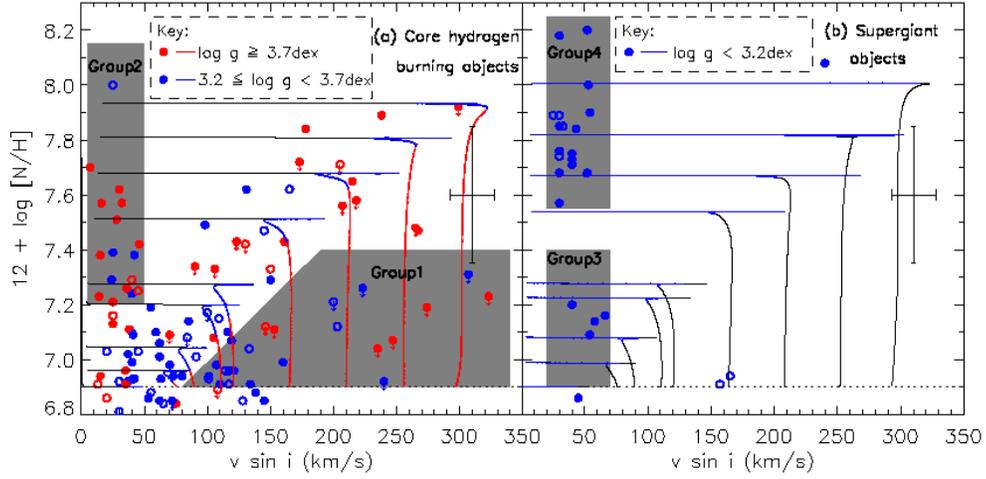

Figure 4: Nitrogen abundances (12+log[N/H]) compared to projected rotational velocities for core-hydrogen-burning (left panel) and supergiant B-type stars (right panel) in the LMC. The solid lines are new evolutionary tracks (Brott et al., in prep.), open circles are radial velocity variables, downward arrows are upper limits to the nitrogen abundance, and the dotted horizontal line is the LMC baseline nitrogen abundance.

Table 1: Present-day composition of the LMC and SMC, taken from Hunter et al. (2007) and Trundle et al. (2007). Abundances are given on the scale 12+log[X/H], with the relative fraction compared to the solar results (Asplund et al., 2005) given in parentheses. Due to uncertainties of the absolute values, the fractions quoted for iron are relative to our Galactic results.

| Element | Solar | LMC | SMC |
| --- | --- | --- | --- |
| C | 8.39 | 7.73 (0.22) | 7.37 (0.10) |
| N | 7.78 | 6.88 (0.13) | 6.50 (0.05) |
| O | 8.66 | 8.35 (0.49) | 7.98 (0.21) |
| Mg | 7.53 | 7.06 (0.34) | 6.72 (0.15) |
| Si | 7.51 | 7.19 (0.48) | 6.79 (0.19) |
| Fe | 7.45 | 7.23 (0.51) | 6.93 (0.27) |

Table 2: Effective temperatures of B-type stars as a function of spectral type, metallicity, and luminosity class, taken from Trundle et al. (2007). The values in parentheses are interpolated.

| Spectral Type | Milky Way | LMC | | | SMC | | |
| --- | --- | --- | --- | --- | --- | --- | --- |
| | V | I | III | V | I | III | V |
| B0 | 30,650 | 28,550 | 29,100 | 31,400 | 27,200 | | 32,000 |
| B0.2 | (29,050) | (26,950) | (27,850) | 30,250 | (25,750) | | 30,800 |
| B0.5 | 27,500 | 25,350 | (26,650) | 29,100 | (24,300) | | 29,650 |
| B0.7 | (25,900) | 23,750 | (25,400) | (27,950) | (22,850) | 25,300 | 28,450 |
| B1 | 24,300 | 22,150 | 24,150 | 26,800 | 22,350 | 23,950 | 27,300 |
| B1.5 | 22,700 | 20,550 | 22,950 | 25,700 | 20,650 | (22,550) | (26,100) |
| B2 | 22,100 | 18,950 | 21,700 | 24,550 | 18,950 | 21,200 | 24,950 |
| B2.5 | 19,550 | 17,350 | 20,450 | 23,400 | 17,200 | 19,850 | |
| B3 | 17,950 | 15,750 | 19,250 | | 15,500 | 18,450 | |
| B5 | | 14,150 | | | 13,800 | | |